\documentclass[a4paper,12pt]{article}
\usepackage{graphicx}
\usepackage{color}
\usepackage{a4wide}

\title{\bf Adaptive multiresolution computations applied to detonations}
\author{
Olivier Roussel$^1\footnote{Corresponding author: rousselo@coria.fr}$ and Kai Schneider$^2$ \\ ~ \\
{\small $^1$CORIA -- UMR 6614 CNRS, Normandie Universit\'e and INSA Rouen}\\ 
{\small Avenue de l'Universit\'e, 76801 Saint-\'Etienne-du-Rouvray Cedex, France.} \\ ~ \\ 
{\small $^2$M2P2 -- UMR 7340 CNRS, Aix-Marseille Universit\'e and Ecole Centrale de Marseille} \\
{\small 38, Rue Fr\'ed\'eric Joliot-Curie, 13451 Marseille Cedex 13, France. } 
}
\date{}
\begin{document}

\maketitle

%-------------------------------------------------------------------
\begin{abstract}
A space-time adaptive method is presented for the reactive Euler equations describing chemically reacting gas flow where a two species model is used for the chemistry.
The governing equations are discretized with a finite volume method and dynamic space adaptivity is introduced
using multiresolution analysis.
A time splitting method of Strang is applied to be able to consider stiff problems while keeping
the method explicit.
For time adaptivity an improved Runge--Kutta--Fehlberg scheme is used.
Applications deal with detonation problems in one and two space dimensions.
A comparison of the adaptive scheme with reference computations on a regular grid allow to assess
the accuracy and the computational efficiency, in terms of CPU time and
memory requirements. 
\end{abstract}
%-------------------------------------------------------------------

% -------------------------------------------------------------------------------------------------------------------------
\section{Introduction}
% -------------------------------------------------------------------------------------------------------------------------

Real world industrial or environmental problems, {\it e.g.}, management of industrial risks, typically involve physical and chemical phenomena having a multitude of dynamically active spatial and temporal scales. 
Their direct numerical modelling thus leads to prohibitive computational cost. 
Introducing adaptivity can be understood in the sense that the computational effort is
concentrated at locations and time instants where it is necessary to ensure a given numerical accuracy, 
while efforts may be significantly reduced elsewhere. 
Adaptive methods are in many cases more competitive than schemes on regular fine grids, in
particular for solutions of nonlinear PDEs exhibiting a non-uniformly distributed regularity of the solution. 
Reliable error estimators of the solution are essential ingredients of fully adaptive schemes.
They are based for example on Richardson ideas of extrapolation, adjoint problems or gradient based approaches. 
For evolutionary problems, a major task is the time evolution of the grid and
its reliable prediction for the next time step. However, to become efficient, adaptive methods require a
significant effort on implementing data structures, which are typically based on graded trees, 
hash-tables or multi-domains.
Moreover, the computational cost per cell is significantly increased with respect to uniform discretizations. Hence, an adaptive method is only efficient when the
data compression is large enough to compensate the additional computational cost per cell. Fortunately, for problems
exhibiting local discontinuities or steep gradients, adaptive computations are faster than fine grid computations.

Adaptive discretization methods for solving nonlinear PDEs have a long tradition and can be tracked back to the late
seventies \cite{Bra77}. Adaptive finite element methods have a long history, in particular for elliptic problems. 
For chemically reactive flow with detailled chemical reaction in three dimensions \cite {BrRi06, BrRi07} proposed stabilized finite elements with adaptive mesh refinement. The equations are treated fully coupled with a Newton solver and the  solution of large linear non-symmetric, indefinite systems becomes necessary for which a parallel multigrid solver is used.
Moving grid techniques have been applied successfully to combustion problems \cite{HLP03}. A posteriori error estimators have also been studied for a long time to improve the grid, since the early work of Babuska and Rheinboldt \cite{BR78}. 
%However, adjoint problems have to be solved, which are more expensive than the original PDEs \cite{BR01}.
However, adjoint problems have to be solved which are linear although the original PDE can be nonlinear \cite{BR01}.
Fully adaptive finite element discretizations of reaction-diffusion problems encountered in electrocardiology have been proposed in \cite{Lang95,FDELP06}. For time adaptivity a stepsize control with linearly implicit time integrators is used. In space a multilevel finite element method is combined with a posteriori local error estimators. 

The main challenge is to estimate and control the error of adaptive schemes with respect to the exact solution, or at least with respect to the
same numerical scheme on an underlying uniform grid. Self adaptive methods are preferred as they automatically adjust to the
solution. The block-structured adaptive mesh refinement technique (AMR or SAMR) for hyperbolic partial differential equations
has been pioneered by Berger and Oliger~\cite{BO84}. While the first approach utilized rotated refinement
grids that required complicated conservative interpolation operations, AMR denotes today especially the simplified variant of
Berger and Collela \cite{BC88} that allows only refinement patches aligned to the coarse grid mesh. The striking efficiency 
of this algorithm, in particular for 3D instationary supersonic gas dynamics problems, has been demonstrated by Berger 
{\it et al.} in \cite{BBS94}.

Recently, multiresolution (MR) techniques have become popular for hyperbolic conservation laws, going back to the seminal work
of Harten \cite{Har95} in the context of finite volume schemes and cell-average MR analysis. Starting point is a finite volume scheme
for hyperbolic conservation laws on a regular grid. Subsequently a discrete multiresolution analysis is used to avoid
expensive flux computations in smooth regions, first without reducing memory requirements, {\it e.g.} for 1D hyperbolic
conservation laws (Harten [30]), 1D conservation laws with viscosity (Bihari \cite{Bih97}), 2D hyperbolic conservations laws (Bihari and Harten \cite{BH96}), 2D compressible Euler equations (Chiavassa and Donat \cite{CD01}), 2D hyperbolic conservation laws with curvilinear
patches (Dahmen {\it et al} \cite{DGM01}) and unstructured meshes (Abgrall and Harten \cite{AH98}, Cohen {\it et al} \cite{CDK00}). A fully adaptive version, 
still in the context of 1D and 2D hyperbolic conservation laws, has been developed to reduce also memory requirements 
(Gottschlich-M\"uller and M\"uller \cite{GM99}, Kaibara and Gomes \cite{KG00}, Cohen {\it et al} \cite{CKM03}). This algorithm has been extended
to the 3D case and to parabolic PDEs (Roussel {\it et al} \cite{RST03}, Roussel and Schneider \cite{RS05}), and more recently to self-adaptive global and local time-steppings (M\"uller and Stiriba \cite{MS07}, Domingues {\it et al} \cite{DGR08,DGR09,DRS09}). Therewith the solution is represented and computed on a dynamically evolving automatically adapted space-time grid. Different strategies have been proposed to evaluate the flux without requiring a full knowledge of fine grid cell-average values.
Applications to shock waves in compressible flows addressing the issue of shock resolution have been presented in \cite{BRT06}, and extensions to the Navier--Stokes equations in the weakly compressible regime can be found in \cite{RS10}.
Adaptive MR methods with operator splitting have been proposed for multiscale reaction fronts
with stiff source terms in \cite{DMDTDLL12,Duar11}.
The numerical analysis of the above higher order operator splitting techniques has been performed in \cite{DDDLM11,DMD11} and it was shown that the splitting time step can be even larger than the times scales involved in the PDEs.
Adaptive MR computations using the above method with complex chemistry and including detailed transport can be found in \cite{DDDLLM14}, further  applications involving various stiffness levels have been presented in \cite{DDTCM13} \cite{DBMBDD12}.

The MR approach has also been used in other contexts. For instance, the Sparse Point Representation (SPR) method was the first
fully adaptive MR scheme, introduced by H\"olmstrom \cite{Hol99} in the context of finite differences and point-value MR analysis,
leading to both CPU time and memory reduction. In the SPR method, the wavelet coefficients are used as regularity indicators
to create locally refined grids, on which the numerical solution is represented and the finite difference discretization of
the operators is performed. Applications of the SPR method have been published in \cite{DGD03,PDF07}. 
Discontinuous Galerkin methods, they have been applied to hyperbolic conservation laws in \cite{CDG05} using Haar wavelet indicators to decide where to refine or to coarsen the meshes. 
These publications reveal that the multiresolution concept has been applied by several groups
with success to different stiff problems. For comprehensive literature about the subject, we refer to the books of Cohen~\cite{Coh00} and M\"uller~\cite{Mue03}.

The objective of the paper is the extension of the adaptive multiresolution method~\cite{RST03,DGR09} to the numerical simulation of detonation waves.
As model we use here a one-step chemical reaction involving two chemical species only.
Since the chemical source term is stiff, a time splitting has to be made between the convective and source terms and each term needs to be computed with a different time step and a different time integration method. An error-controled time step based on a Runge--Kutta--Fehlberg approach is used 
for the source term integration.
The application concerns Chapman-Jouguet detonations in one dimension with different stiffness values and instabilities of detonations waves due to an interaction with a pocket of partially burnt gases in two dimensions.
{These detonation problems motivate the use of  the Euler equations, although extensions of the method to compute viscous flows using the Navier--Stokes equations is straightforward, as proposed in \cite{RS10}.}

The outline of the paper is the following: first, we present the set of reactive Euler equations for a simplified detonation
model. Then we describe the finite volume discretization. A Strang splitting technique is utilized to account for
temporal scales in the source term that do not influence the hydrodynamics. 
We also briefly summarize the multiresolution strategy. 
Finally, we show the numerical results for the test problems in one and two space dimensions and we set the conclusions together with some perspectives for future work.

% -------------------------------------------------------------------------------------------------------------------------
\section{Governing equations}
% -------------------------------------------------------------------------------------------------------------------------

For modelling the combustion process, we use the reactive Euler equations, as described in \cite{CF85,GR95}. 
The simplest description of a chemically reacting gas flow assumes that the gas mixture is made only of two chemical species, the burnt
gas, denoted with subscript $b$ and the unburnt gas, denoted with subscript $u$. The unburnt gas is converted to burnt gas
via a single irreversible reaction. We represent the mixture state by a single scalar variable $Z$ corresponding the mass
fraction of the unburnt gas. We also assume gases in the mixture to be ideal polytropic gases with equal specific heat ratio
$\gamma$ and specific gas constant $r$. 
The system of equations in two dimensions, { which has been non-dimensionalized in a suitable way,} may be written as

\begin{equation} \label{eqn:pde}
\frac{\partial Q}{\partial t} + \frac{\partial F}{\partial x} +  \frac{\partial G}{\partial y} = S,
\end{equation}

\bigskip \noindent where $Q = (\rho, \rho v_x, \rho v_y, \rho e, \rho Z)^T$ and

\begin{equation}
F = \left(
\begin{array}{c}
\rho v_x \\ \rho v_x^2+p \\ \rho v_x v_y \\ (\rho e+p)v_x \\ \rho v_x Z
\end{array}
\right) \, , \,
G = \left(
\begin{array}{c}
\rho v_y \\ \rho v_x v_y \\ \rho v_y^2+p \\ (\rho e+p)v_y \\ \rho v_y Z
\end{array}
\right) \, , \,
S = \left(
\begin{array}{c}
0 \\ 0 \\ 0 \\ 0 \\ -K(T) \rho Z
\end{array}
\right)
\end{equation}

\bigskip \noindent Here $\rho$ denotes the mixture density, $V = (v_x,v_y)^T$ the mixture velocity, $e$ the mixture total
energy per unit of mass, $p$ the pressure, $T$ the temperature and $k$ the chemical reaction rate. The two equations of state completing the model are

\begin{equation}
p = \rho r T
\end{equation}

\bigskip \noindent and

\begin{equation}
e = \frac{p}{\rho(\gamma -1)} + \frac{V^2}{2} + Q_0 Z
\end{equation}

\bigskip \noindent where $Q_0$ denotes the amount of heat per unit of mass released in the chemical reaction.

The reaction rate $k(T)$ of the irreversible chemical reaction is expressed in Arrhenius form as

\begin{equation}
k(T) = A \exp \left( -\frac{T_A}{T} \right)
\end{equation}

\bigskip \noindent where the pre-exponential coefficient $A$ and the activation temperature $T_A$ are empirical constants.
When the reaction source term is stiff, however, the reaction rate may be simplified by adopting the so-called ignition
temperature kinetic model, {\it i.e.}

\begin{equation}
k(T) = \left\{
\begin{array}{lll}
\frac{1}{\tau} & \mbox{ if } & T \geq T_i \\
0 & \mbox{ if } & T < T_i
\end{array}
\right.
\end{equation}

\bigskip \noindent where $T_i$ denotes the ignition temperature and $\tau$ the characteristic time of the chemical reaction,
which determines the stiffness of the problem.
{ This formulation has been chosen in the applications presented in the numerical results section. 
However, the numerical method is not limited to this simplified model and its extension to the Arrhenius law or even more complex chemical reactions is possible.}

% -------------------------------------------------------------------------------------------------------------------------
\section{Numerical method}
% -------------------------------------------------------------------------------------------------------------------------
{ In this section, we first describe the classical Strang splitting and the space discretization of the convective terms.
Subsequently, the time integration is discussed and, first, for the convective terms, a time step depending on the CFL
is chosen, then, for the source term, we split the convective time step into error-controlled time steps using an explicit Dormand-Prince
method. A different choice has been proposed by Duarte {\it et al } in \cite{DMDTDLL12}, based on implicit and explicit Runge-Kutta methods
and a posteriori error estimators. We also briefly recall the multiresolution method, previously published in \cite{RST03,DGR08}.} 
% -------------------------------------------------------------------------------------------------------------------------
\subsection{Strang splitting}

We denote by $C(Q)$ the operator of the convective terms. Equation (\ref{eqn:pde})
becomes
\begin{equation}
\frac{\partial Q}{\partial t} = C(Q) + S(Q).
\end{equation}

\bigskip \noindent Discretizing explicitly with first order in time, we get
\begin{equation}
Q^{n+1} = Q^{n} + \Delta t \left[ C(Q^n) + S(Q^n) \right]
\end{equation}

\bigskip \noindent where $n$ denotes the time instant and $\Delta t$ the convective time step.

The splitting relies on the separation of the convective and source terms operators. To get a first-order
discretization, it writes
\begin{eqnarray}
Q^{\star} & = & Q^{n} + \Delta t \; S(Q^n) \\
Q^{n+1} & = & Q^{\star} + \Delta t \; C(Q^{\star})
\end{eqnarray}

\bigskip \noindent {\it i.e.}
\begin{equation}
Q^{n+1} =  C^{o1}_{\Delta t} \; S^{o1}_{\Delta t} \; Q^n
\end{equation}

\bigskip \noindent where $S^{o1}_{\Delta t}$ denotes the first-order accurate source term operator with time
step $\Delta t$ and $C^{o1}_{\Delta t}$ the first-order accurate convection operator with time step $\Delta t$.

\bigskip \noindent To get second-order accuracy, { which corresponds to the classical Strang,} the following procedure can be applied
\begin{equation}
Q^{n+1} = S^{o2}_{\Delta t/2} \; C^{o2}_{\Delta t} \; S^{o2}_{\Delta t/2} \; Q^n
\end{equation}

\bigskip \noindent where $S^{o2}_{\Delta t/2}$ denotes the second-order accurate source term operator with time
step $\Delta t/2$ and $C^{o2}_{\Delta t}$ the second-order accurate convection operator with time step $\Delta t$.

{ We note that the splitting time step in the above method is fixed. Techniques to introduce adaptive splitting time steps
based on a posteriori error estimators have been introduced in \cite{DDDLM11,Duar11} allowing automatic error control.}

% -------------------------------------------------------------------------------------------------------------------------
\subsection{Space discretization of the convective terms}

Discretization in space is made using a finite volume method, to ensure conservative flux computations.
Convective terms are discretized using the AUSM+ scheme \cite{Lio96}. In this procedure,
pressure terms are computed separately. The Euler flux $F$ writes
\begin{equation}
F = \left(
\begin{array}{c}
\rho v_x \\ \rho v_x^2+p \\ \rho v_x v_y \\ (\rho e+p)v_x \\ \rho v_x Z
\end{array}
\right) = M \; c \; \left(
\begin{array}{c}
\rho \\ \rho v_x \\ \rho v_y \\ \rho h \\ \rho  Z
\end{array}
\right) + \left(
\begin{array}{c}
0 \\ p \\ 0 \\ 0 \\ 0
\end{array}
\right) 
\end{equation} 

\bigskip \noindent where $M$ denotes the Mach number, $c$ the speed of sound and $h$ the enthalpy per unit of mass. 
Denoting by $\Phi$ the purely convective
term and by $\Pi$ the pressure term, we discretize the Euler flux $F$ in space following
\begin{equation}
F_{i+\frac{1}{2}} = M_{i+\frac{1}{2}} \; c_{i+\frac{1}{2}} \; \Phi_{i+\frac{1}{2}} + \Pi_{i+\frac{1}{2}}
\end{equation}
{ where the indices refer to nodal values.}

\bigskip \noindent The interface speed of sound is $c_{i+\frac{1}{2}} = \sqrt{c_{i} c_{i+1}}$ and the interface convective
term is
\begin{equation}
\Phi_{i+\frac{1}{2}} = \left\{
\begin{array}{ll}
\Phi_i & \mbox{ if } M_{i+\frac{1}{2}} \geq 0 \\
\Phi_{i+1} & \mbox{ otherwise}
\end{array}
\right.
\end{equation}

\bigskip \noindent The terms $M_{i+\frac{1}{2}}$ and $p_{i+\frac{1}{2}}$ follow
\begin{eqnarray}
M_{i+\frac{1}{2}} & = & M^+_i + M^-_{i+1} \\
p_{i+\frac{1}{2}} & = & P^+_i \; p_i + P^-_{i+1} \; p_{i+1}
\end{eqnarray}

\bigskip \noindent where
\begin{equation}
M_i^\pm = \left\{
\begin{array}{ll}
\frac{1}{2} \left( M_i \pm |M_i| \right) & \mbox{ if } |M_i| \geq 1 \\
\pm \frac{1}{2} \left(M_i \pm 1 \right)^2 \pm \frac{1}{8} \left( M_i^2 - 1 \right)^2 & \mbox{ otherwise}
\end{array}
\right.
\end{equation}

\bigskip \noindent and
\begin{equation}
P_i^\pm = \left\{
\begin{array}{ll}
\frac{1}{2} \left( 1 \pm \mbox{sign}(M_i) \right) & \mbox{ if } |M_i| \geq 1 \\
\frac{1}{4} \left(M_i \pm 1 \right)^2\left( 2 \mp M_i \right) 
\pm \frac{3}{16} M_i \left( M_i^2 - 1 \right)^2 & \mbox{ otherwise}
\end{array}
\right.
\end{equation}

\bigskip
Third-order accuracy {of the spatial discretization} far from discontinuities is obtained using a MUSCL interpolation \cite{Van79}, together with a 
Koren slope limiter \cite{Kor93}. Denoting by $q$ one of the conservative quantities ({\it i.e.} density, momentum, energy, 
partial mass of the unburnt gas), the corrected value via a MUSCL interpolation is
\begin{equation}
q'_{i} = q_i + \frac{1}{6}\phi \left( r_i \right) \left(q_i-q_{i-1} \right) + \frac{1}{3} \phi \left( \frac{1}{r_i} \right)
\left(q_{i+1} - q_{i} \right) 
\end{equation}

\bigskip \noindent and
\begin{equation}
q'_{i+1} = q_{i+1} - \frac{1}{3}\phi \left( r_{i+1} \right) \left(q_{i+1}-q_i \right) - \frac{1}{6} \phi \left( 
\frac{1}{r_{i+1}} \right) \left(q_{i+2} - q_{i+1} \right) 
\end{equation}

\bigskip \noindent where
\begin{equation}
r_i =\frac{q_{i+1}-q_i}{q_i-q_{i-1}}
\end{equation}

\bigskip \noindent and
\begin{equation}
\phi(r) = \max \left[ 0, \min \left( 2r, \frac{1+2r}{3},2\right)\right]
\end{equation}

\bigskip \noindent Nevertheless, since we use a second-order accurate Strang splitting {in time}, the global accuracy of the scheme remains second-order.

% -------------------------------------------------------------------------------------------------------------------------
\subsection{Time integration of the convective terms}

Time integration of the convective term is made using a classical third-order TVD Runge-Kutta scheme, i.e.
\begin{eqnarray}
Q^{\star} & =  & Q^n + \Delta t \; C\left( Q^n \right) \\
Q^{\star\star} & = & \frac{1}{4} \left[ 3 \; Q^n + Q^\star + \Delta t \; C \left( Q^\star \right) \right] \\
Q^{n+1} & = & \frac{1}{3} \left[ Q^n + 2 \; Q^{\star\star} + 2 \; \Delta t \; C \left( Q^{\star\star} \right) \right]
\end{eqnarray}

\bigskip \noindent The corresponding Runge-Kutta tableau is given in Table \ref{table:RK3}.

\begin{table}[htbp]
\begin{center}
\begin{tabular}{c|ccc}
$0$ & ~ & ~ & ~ \\
& & & \\
$1$ & $1$ & ~ & ~ \\
& & & \\
$\frac{1}{4}$ & $\frac{1}{4}$ & $\frac{1}{4}$ & $0$ \\
& & & \\
\hline
& & & \\
~ & $\frac{1}{6}$ & $\frac{1}{6}$ & $\frac{2}{3}$  
\end{tabular}
\caption{Butcher tableau corresponding to the compact TVD third-order Runge-Kutta method.}
\label{table:RK3}
\end{center}
\end{table}

\bigskip The convective time step $\Delta t$ is chosen to satisfy the Courant-Friedrichs-Levy (CFL) condition. In the
computations, $CFL = 0.5$ is used.

% -------------------------------------------------------------------------------------------------------------------------
\subsection{Time integration of the stiff source terms}

Due to the stiffness of the chemical source terms in case of detonation computations, a high order time integration is chosen to  ensure the numerical stability. 
{ As the chemical time scale is much smaller than the convective one
we apply a much smaller time step $\Delta t_d$ than for the convection terms. }
However, this is only required in a small area of the computational domain, {\it i.e.} in the
reaction zone. In the rest of the computational domain, the source term is almost equal to zero.

For this reason, we introduce an adaptive local time step $\Delta t_d = \frac{\Delta t}{2N}$, $N$ being the number of substeps
required by the source term computation. In order to adapt this local time step with time, a fourth-fifth order embedded
Runge-Kutta method is used. The computational error between the fourth-order and the fifth-order method enables to decide
wether the local time step needs to be increased or decreased, as proposed by Fehlberg \cite{Feh64}.

\begin{eqnarray}
Q^{n+1}_{o4} & = & Q^n + \Delta t_d \; \sum_{i=1}^{s} b_{i,o4} k_i \\
Q^{n+1}_{o5} & = & Q^n + \Delta t_d \; \sum_{i=1}^{s} b_{i,o5} k_i \\ 
\end{eqnarray}

\bigskip \noindent where
\begin{equation}
k_i = S \left( t^n + c_i \Delta t_d, Q^n + \Delta t_d \; \sum_{i=1}^{s} \sum_{j=1}^{s} a_{i,j} k_j \right)
\end{equation}

\bigskip \noindent In this article, an embedded explicit Dormand-Prince method~\cite{DP80} was chosen. The coefficients are
computed in order to minimize the error of the fifth-order solution. This is the main difference with the Fehlberg method,
which was constructed so that the fourth-order solution has a small error. For this reason, the Dormand-Prince method is 
more suitable when the higher-order solution is used to continue the time integration. The coefficients are given in 
Table \ref{table:DP}.

\begin{table}[htbp]
\begin{center}
\begin{tabular}{c|ccccccc}
$0$ & & & & & & & \\
& & & & & & & \\
$\frac{1}{5}$ & $\frac{1}{5}$ & & & & & & \\
& & & & & & & \\
$\frac{3}{10}$ & $\frac{3}{40}$ & $\frac{9}{40}$ & & & & & \\ 
& & & & & & & \\
$\frac{4}{5}$ & $\frac{44}{45}$ & $-\frac{56}{15}$ & $\frac{32}{9}$ & & & & \\
& & & & & & & \\
$\frac{8}{9}$ & $\frac{19372}{6561}$ & $-\frac{25360}{2187}$ & $\frac{64448}{6561}$ & $-\frac{212}{729}$ & & & \\
& & & & & & & \\
$1$ &	$\frac{9017}{3168}$ & $-\frac{355}{33}$ & $\frac{46732}{5247}$ & $\frac{49}{176}$ & $-\frac{5103}{18656}$ & & \\
& & & & & & & \\
$1$ &	$\frac{35}{384}$ & $0$ & $\frac{500}{1113}$ & $\frac{125}{192}$ & $-\frac{2187}{6784}$ & $\frac{11}{84}$ & \\ 	
& & & & & & & \\
\hline
& & & & & & & \\
& $\frac{5179}{57600}$ & $0$ & 	$\frac{7571}{16695}$ & 	$\frac{393}{640}$ & $-\frac{92097}{339200}$ & $\frac{187}{2100}$ 
& $\frac{1}{40}$ \\
& & & & & & & \\
& $\frac{35}{384}$ & $0$ & $\frac{500}{1113}$ & $\frac{125}{192}$ & $-\frac{2187}{6784}$ & $\frac{11}{84}$ & $0$
\end{tabular}
\caption{Butcher tableau corresponding to the Dormand-Prince method. The first row of $b$ coefficients gives the fourth-order accurate solution, and the second row yields order five.}
\label{table:DP}
\end{center}
\end{table}

The local relative error of the quantity $q$ is denoted by $e$ and the accepted tolerance by $\varepsilon$. The optimal
time step $\sigma \; \Delta t_d$ can be determined by multiplying the scalar $\sigma$ times the current time step $\Delta t_d$.
The scalar $\sigma$ is given by
\begin{equation}
\sigma = \left( \frac{\varepsilon \; \Delta t_d}{2 \; e \; \Delta t_{max}} \right) ^{\frac{1}{4}} =
\left( \frac{\varepsilon \; \Delta t_d}{ e \; \Delta t} \right) ^{\frac{1}{4}}
\end{equation}

In practice, however, the new time step $\Delta t'_d$ is determined in function of $\sigma$, following
\begin{equation}
\Delta t'_d = \left\{
\begin{array}{ll}
2 \; \Delta t_d & \mbox{ if } \sigma > 2 \\
\frac{\Delta t_d}{2} & \mbox{ if } \sigma < 1 \\
\Delta t_d & \mbox{ otherwise}
\end{array}
\right.
\end{equation}

% -------------------------------------------------------------------------------------------------------------------------
\subsection{Multiresolution method} %kai: Olivier please check again to avoid to much-copy paste w.r.t. RSTB03

The principle in the multiresolution (MR) setting is to represent a set of function
cell averages as values on a coarser grid plus a series of differences
at different levels of nested grids. 
The differences contain the information
of the function when going from a coarse to a finer grid. 

A tree data structure is  an efficient way to store the reduced MR dataas it allows to 
reduce the memory with respect to a finite volume (FV) scheme on the finest level. 
This representation  could also  increase the speed-up during the time evolution because it reduces the time-searching of the elements. 

\begin{figure}[htbp]
\begin{center}
\includegraphics[scale=0.4]{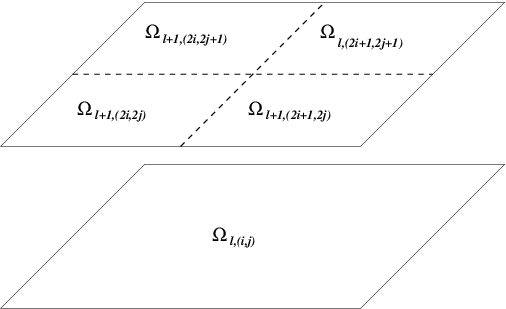}
\end{center}
\caption{\label{cap:Dyadic-refinement}Dyadic grid refinement in 2D.}
\end{figure}

In the following we consider a hierarchy of regular grids in 2D, $\Omega_{\ell}$, $0\leq\ell\leq L$.
The root cell is $\Omega_{0,0,0}=\Omega$ and corresponds to a rectangle
with side lengths $h_{x}$ and $h_{y}$. 
The different node cells at a level $\ell>0$ forming $\Omega_{\ell}$ are denoted by  
$\Omega_{\ell,i,j}$ where  $(i,j)\in\Lambda_{\ell}$. 
The ensemble of indices of the existing node cells on the level $\ell$ is $\Lambda_{\ell}$.  
Note that $\Omega_{\ell,i,j}$ are rectangles with side lengths
$h_{x,\ell}=2^{-\ell}h_{x}$ and $h_{y,\ell}=2^{-\ell}h_{y}$.
In the tree terminology, the refinement of a parent node cell $\Omega_{\ell,i,j}$ at level
$\ell$ produces four children nodes $\Omega_{\ell+1,2i,2j}$,
$\Omega_{\ell+1,2i,2j+1}$, $\Omega_{\ell+1,2i+1,2j}$ and $\Omega_{\ell+1,2i+1,2j+1}$
at level $\ell+1$, as illustrated in Figure \ref{cap:Dyadic-refinement}.

%--------------------------------------------------------------------------------------------------
%--------------------------------------------------------------------------------------------------

The cell-average value of the quantity $u$ on the cell $\Omega_{\ell,i,j}$ is given by
$\bar{u}_{\ell,i,j}=\frac{1}{|\Omega_{\ell,i,j}|}\int_{\Omega_{\ell,i,j}}u(x,y) \; dx \; dy$.
and correspondingly the ensemble of the existing cell-average values at level $\ell$ by
$\bar{U}_{\ell}=(\bar{u}_{\ell,i,j})_{(i,j)\in\Lambda_{\ell}}$.
The projection { (or restriction)} operator 
\[
P_{\ell+1\rightarrow l}\,:\, \bar{U}_{\ell+1}\,\mapsto\,\bar{U}_{\ell}.
\]
estimates the cell-averages of a level $\ell$ from the ones of
the level $\ell+1$.
The parent cell-average is the weighted average
of the children cell-averages
\[
\bar{u}_{\ell,i,j}=\frac{1}{4}(\bar{u}_{\ell+1,2i,2j}+\bar{u}_{\ell+1,2i,2j+1}+
\bar{u}_{\ell+1,2i+1,2j}+\bar{u}_{\ell+1,2i+1,2j+1})
\]
and  thus the projection operator is exact and unique.
To predict the cell-averages of a level $\ell+1$ from the ones of
the level $\ell$, we use a prediction { (or interpolation)} operator 
\[
P_{\ell\rightarrow l+1}\,:\, \bar{U_{\ell}}\,\mapsto\,\widetilde{U}_{\ell+1}.
\]
 
\bigskip \noindent This operator yields an approximation $\widetilde{U}_{\ell+1}$
of $\bar{U}_{\ell+1}$ at the level $\ell+1$. 
In this paper, we use third order interpolation given by a tensor product approach 
\cite{BH96}.
For $n,p\in\{0,1\}$, we define
\begin{eqnarray}
\widetilde{u}_{\ell+1,2i+n,2j+p} & = & \bar{u}_{\ell, i,j} + \frac{1}{8}(-1)^{n}
\left( \bar{u}_{\ell,i+1,j}-\bar{u}_{\ell,i-1,j} \right) \nonumber \\
 & & +  \frac{1}{8}(-1)^{p} \left( \bar{u}_{\ell,i,j+1}-\bar{u}_{\ell,i,j-1} \right) \\
 & & +\frac{1}{64}(-1)^{np} \left[ \left( \bar{u}_{\ell,i+1,j+1}-\bar{u}_{\ell,i+1,j-1} \right) -
 \left( \bar{u}_{\ell,i-1,j+1}-\bar{u}_{\ell,i-1,j-1} \right) \right]. \nonumber
\end{eqnarray}
First, this prediction is local, since it is made from the cell
average $\bar{u}_{\ell,i,j}$ and the eight nearest uncles $\bar{u}_{\ell,i\pm1,j\pm1}$.
Second, it is consistent with the projection, {\it i.e.}
$P_{\ell+1\rightarrow\ell}\circ P_{\ell\rightarrow\ell+1}=\mbox{Id}$.
%
%Defining wavelet coefficients as the difference between the exact and the predicted 
%values  three children cells
%
The difference between the exact and the predicted values at three children cells yields
the wavelet (or detail) coefficients.
The sum of the four details in the children cells
is equal to zero \cite{BH96} 
\begin{eqnarray}
\bar{d}_{\ell+1,2i,2j+1} & = &  \bar{u}_{\ell+1,2i,2j+1}-\widetilde{u}_{\ell+1,2i,2j+1} \nonumber \\
\bar{d}_{\ell+1,2i+1,2j} & = &  \bar{u}_{\ell+1,2i+1,2j}-\widetilde{u}_{\ell+1,2i+1,2j}\\
\bar{d}_{\ell+1,2i+1,2j+1} & = & \bar{u}_{\ell+1,2i+1,2j+1}-\widetilde{u}_{\ell+1,2i+1,2j+1}. \nonumber
\end{eqnarray}

%it results that the knowledge of the cell-average values on the children
This implies that the knowledge of the cell-average values on the children
$\bar{U}_{\ell+1}$ is equivalent to the knowledge of the cell-average
values on the parents $\bar{U}_{\ell}$ and the wavelet coefficients
$\bar{D}_{\ell+1}=(\bar{d}_{\ell+1,2i,2j+1},\bar{d}_{\ell+1,2i+1,2j},
\bar{d}_{\ell+1,2i+1,2j+1})_{(i,j)\in\Lambda_{\ell}}$.
The so-called multiresolution transform on the cell-average values
is obtained by repeating this operation recursively on $L$ levels~\cite{Har95},
\[
\bar{U}_{L}\longleftrightarrow(\bar{D}_{L},\,\bar{D}_{L-1},\,\ldots,\bar{D}_{1},\,\bar{U}_{0}).
\]
In conclusion, knowing the cell-average values of all the
leaves $\bar{U}_{L}$ is equivalent to knowing the cell-average
value of the root $\bar{U}_{0}$ and the details of all the other
nodes of the tree structure. 

In the MR scheme, instead of using the representation on the full
uniform grid $\Omega_{L},$ the numerical solution $\bar{U}_{MR}^{n}=\bar{U}_{L,MR}^{n}$
is formed by cell averages on an adaptive sparse grid $\Gamma^{n}=\Gamma_{L}^{n}$.
Grid adaptivity in the MR scheme is related with an incomplete tree structure,
where cell refinement may be interrupted at intermediate scale levels.
This means that $\Gamma^{n}$ is formed by \textit{leaf cells} $\Omega_{\ell,i,j}$, 
$0\leq\ell\leq L$, $(i,j) \in {\cal L}(\Lambda_{\ell})$, which are cells without children.
Here ${\cal L}(\Lambda_{\ell})$ denotes the ensemble of indices for the existing leaf cells
of the level $\ell$.

Three basic steps are undertaken to evolve the solution  from $\bar{U}_{MR}^{n}$ to $\bar{U}_{MR}^{n+1}$,

\bigskip \noindent \textbf{Refinement:} ${\bar{U}}_{MR}^{n+}\leftarrow{\mathbf{R}}\bar{U}_{MR}^{n}$

\bigskip \noindent \textbf{Evolution:} $\check{\bar{U}}_{MR}^{n+1}\leftarrow\mathbf{E}_{MR}{\bar{U}}_{MR}^{n+}$

\bigskip \noindent \textbf{Coarsening:} $\bar{U}_{MR}^{n+1}\leftarrow\mathbf{T}(\epsilon)\check{\bar{U}}_{MR}^{n+1}$

\bigskip \noindent The refinement operator ${\mathbf{R}}$ is a precautionary measure
to account for possible translation or creation of finer scales in
the solution between two subsequent time steps. 
As the regions of smoothness or irregularities of the solution may change with time,
the grid $\Gamma^{n}$ may not be convenient anymore at the next time
step $t^{n+1}$. 
Hence, before doing the time evolution, the representation
of the solution should be extended onto a grid ${\Gamma}^{n+}$, which
is expected to be a refinement of $\Gamma^{n}$, and to contain $\Gamma^{n+1}$.
Then, the time evolution operator $\mathbf{E}_{MR}=\mathbf{E}_{MR}(\Delta t)$
is applied. 
The subscript MR in $E_{MR}$ means that only the cell-averages
on the leaves of the computational grid ${\Gamma}^{n+}$ are evolved in time, and
that an adaptive flux computation $F_{MR}({\bar{U}}_{MR}^{n+})$ is
adopted at interfaces of cells of different scale levels. 
Finally, a thresholding operation $\mathbf{T}(\epsilon)$ (coarsening)
is applied in order to unrefine those cells in ${\Gamma}^{n+}$ that
are unnecessary for an accurate representation of $\bar{U}_{MR}^{n+1}$. 
{ The choice of the threshold value is motivated  by an error analysis equilibrating the perturbation and discretization errors. For details we refer to \cite{CKM03,RST03}.}

To compress data in an adaptive tree structure, while
still being able to navigate through it, \textit{gradedness} is required.
For instance, for a given node in the dynamic tree structure we 
assume that:

\begin{itemize}
\item its parent and eight nearest uncles are in the tree (if not, we create
them as nodes);
\item for flux computations, if $\Omega_{\ell,i,j}$ is a leaf, its four
nearest cousins $\Omega_{\ell,i\pm2,j}$ and $\Omega_{\ell,i,j\pm2}$
in each direction are in the tree (if not, we create them as virtual
leaves);
\item if a child is created, all its brothers are also created;
\end{itemize}
For more details of these procedures, we refer to \cite{RST03}. 

In the tree structure, the thresholding operator $\mathbf{T}(\epsilon)$
is defined by removing leaves where details are smaller than a prescribed
tolerance $\epsilon$, while preserving the gradedness property, and
the refinement  operation ${\mathbf{R}}$ adds one more level as security
zone, in order to forecast the evolution of the solution in the tree
representation at the next time step. 
These two operations are performed by the following procedure.

\begin{figure}[htbp]
\begin{center}\includegraphics[scale=0.4]{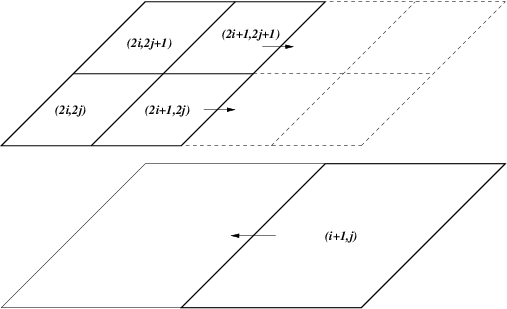}
\end{center}
\caption{\label{cap:Adptive-numerical-flux}Adaptive numerical flux computation in 2D.}
\end{figure}

We denote by $\Lambda$ the ensemble of indices of the existing tree
nodes in $\Gamma^{n+}$, by $\mathcal{L}({\Lambda})$ the restriction
of $\Lambda$ onto the leaves, and by $\Lambda_{\ell}$ the restriction
of $\Lambda$ to a level $\ell$, $0\leq\ell<L$. 
For the whole tree, from the leaves to the root: 

\begin{itemize}
\item Compute the details on the nodes $\bar{d}_{\ell,i,j},\;(i,j)\in\Lambda_{\ell-1}$,
by the multiresolution transform; 

\item Define the deletable cells, if the details on the corresponding nodes
and their brothers are smaller than the prescribed tolerance. 

\end{itemize}
For the whole tree, from the leaves to the root: 

\begin{itemize}
\item If a node and its children nodes are deletable, and the children nodes
are simple leaves (without virtual children), then delete their children. 
\item If the node and its parents are not deletable, and it is not at the maximum level,
then create the children for this node. 
\end{itemize}
To illustrate the adaptive flux computation, we consider the leaf $\Omega_{\ell+1,2i+1,2j}$,
sharing an interface with another leaf $\Omega_{\ell,i+1,j}$ at
a lower scale level, as illustrated in Figure \ref{cap:Adptive-numerical-flux}.
For the calculation of the outgoing numerical flux on the right
interface, we use the cell width in the $x$ direction $h_{x,\ell+1}$
as step size. 
The required right neighboring stencils are obtained from
the cousins $\Omega_{\ell+1,2i+2,2j}$ and $\Omega_{\ell+1,2i+3,2j+1}$,
which are virtual cells. 
For conservation, the ingoing flux on the leaf
$\Omega_{\ell,i+1,j}$ is set equal to the sum of the outgoing fluxes
on the neighbour leaves of level $\ell+1$. 
For more details on the implementation of this procedure we refer
to \cite{RST03}. 

%--------------------------------------------------------------------------------------------------
%--------------------------------------------------------------------------------------------------

% -------------------------------------------------------------------------------------------------------------------------
\section{Numerical results}
% -------------------------------------------------------------------------------------------------------------------------

In all numerical simulations, we consider an { initially} planar Chapman-Jouguet detonation moving with constant unitary speed through
the unburnt gas to the right of the domain. Setting $\gamma = 1.4$, $r=1$, $Q_0=1$ and $T_i=0.22$ and assigning the following
burnt gas values

\begin{equation}
\rho_b = \gamma \; , \, v_{x,b} = \delta \; , \, p_b = 1 \; , Z_b = 0 \, ,
\end{equation}

\bigskip \noindent where
\begin{equation}
\delta = \sqrt{\frac{2(\gamma-1)}{\gamma+1}} \, ,
\end{equation}

\bigskip \noindent
the corresponding von Neumann state past the shock wave is \cite{TV08,HLW00}

\begin{equation}
\rho_N = \frac{\gamma}{1-\delta} \; , \, v_{x,N} = 2\delta \; , \, p_N = 1 + \gamma \delta \; , Z_N = 1 \, ,
\end{equation}

\bigskip \noindent
and the unburnt gas values are

\begin{equation}
\rho_u = \frac{\gamma}{1+\delta} \; , \, v_{x,u} = 0 \; , \, p_u = 1 - \gamma \delta \; , Z_u = 1 \, .
\end{equation}

The resulting temperature of the unburnt gas $T_u=0.215995$ is only slightly lower than the chosen ignition temperature $T_i$.
The initial condition for an initial front at $x=x_0$ is

\begin{equation}
q (x,0) = 
\left\{
\begin{array}{lll}
(q_N - q_b) \exp \left[ a (x-x_0) \right] + q_B & \mbox{ if } & x \leq x_0 \\
q_u & \mbox{ if } & x > x_0
\end{array}
\right.
\end{equation}

\bigskip \noindent where $a$ is set to $\frac{1}{\tau}$. Here $q$ stands for $\rho$, $v_x$, $p$, and $Z$, 
while $v_y=0$ everywhere. This way the initial condition is close to the classical Zeldovich-von Neumann-Dring (ZND) solution of the reaction Euler equations.

% -------------------------------------------------------------------------------------------------------------------------
\subsection{One-dimensional detonation}

First we consider a one-dimensional setting.
The computational domain is $\Omega=[-3,1]$.
In this case, the location of the initial front is set to $x_0 = -2$. 
In order to follow the detonation wave and perform longer computations, a change of variables is made for $v_x$. 
We set $v'_x = v_x - \delta$ and omit the superscript $'$ everywhere. 
Since the initial transition happens in the segment $[-3,-1]$, we only focus on the domain $[-1,1]$. 
The dimensionless final time is $t=2.2$. 

% -------------------------------------------------------------------------------------------------------------------------
\subsubsection{Non-stiff case}

\begin{figure}[htbp]
\includegraphics[width=0.5 \textwidth]{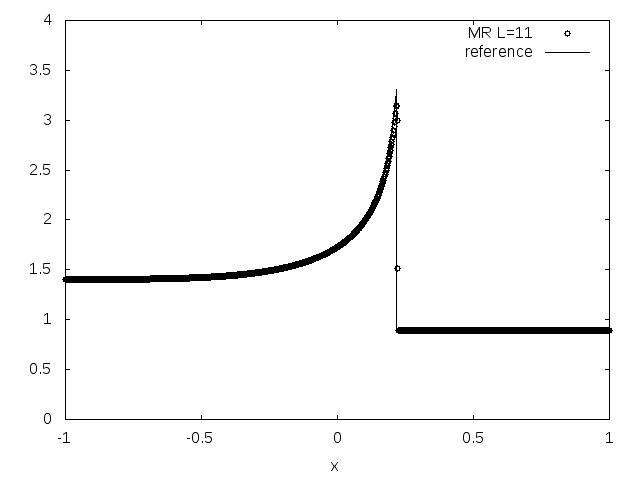}
\includegraphics[width=0.5 \textwidth]{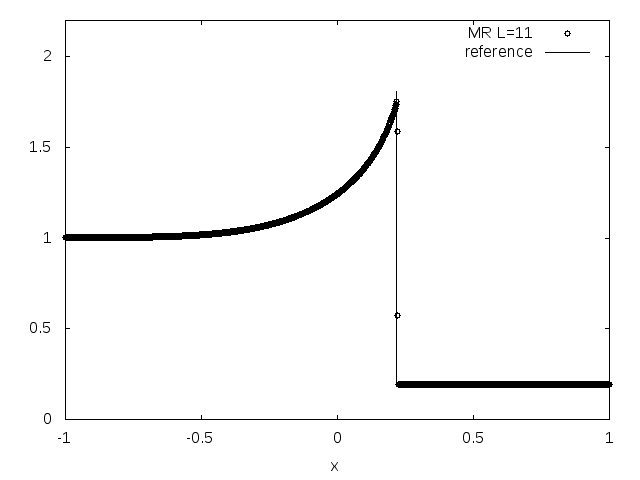}
\\
\includegraphics[width=0.5 \textwidth]{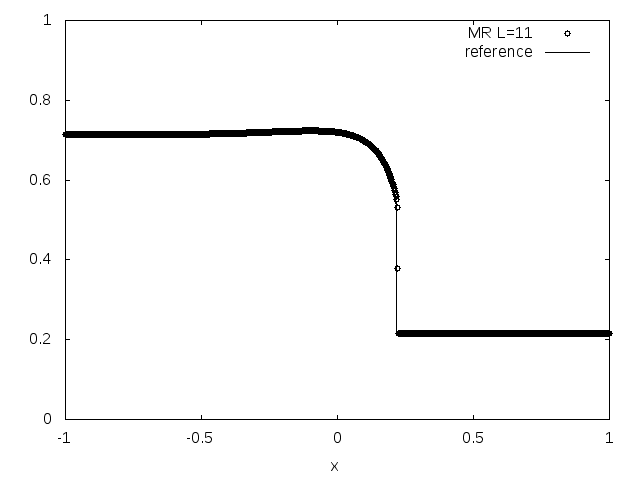}
\includegraphics[width=0.5 \textwidth]{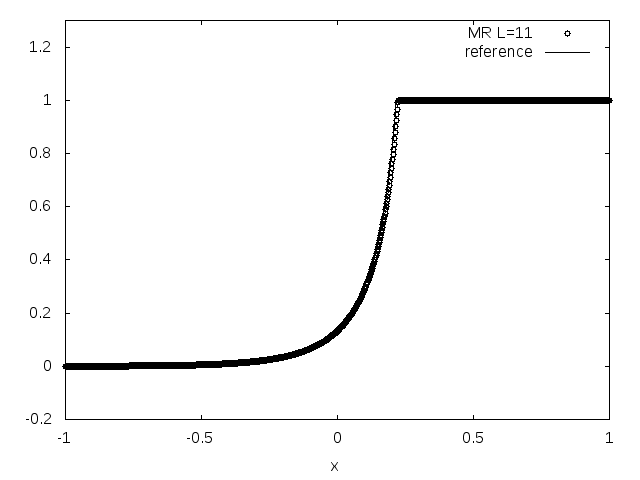}
\\
\includegraphics[width=0.5 \textwidth]{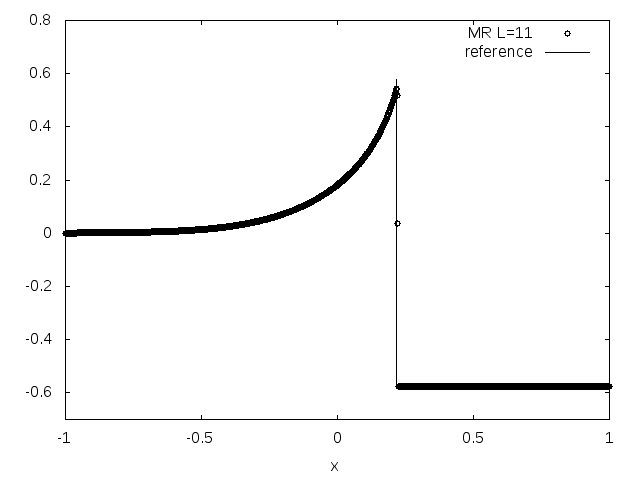}
\includegraphics[width=0.5 \textwidth]{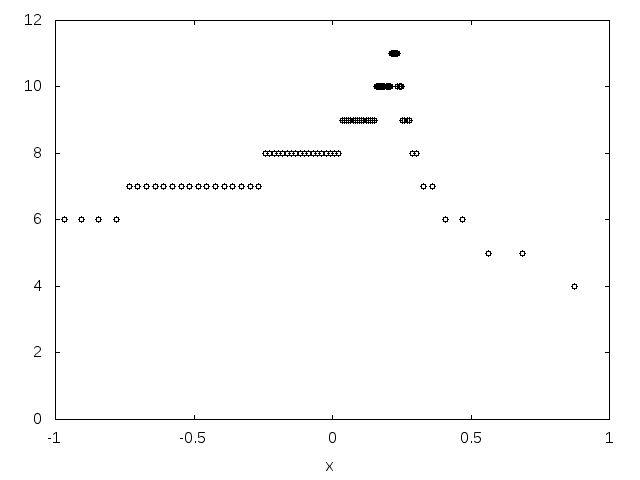}
\caption{Chapman-Jouguet detonation front:
density ({\it top, left}), pressure ({\it top, right}), temperature ({\it middle, left}),
partial mass of the limiting reactant ({\it middle, right}), velocity ({\it bottom, left}) and 
adaptive mesh ({\it bottom, right}) at $t=2.2$, $L=11$ scales, $\epsilon=2.5 \cdot 10^{-3}$, $\tau=10^{-1}$.}
\label{fig:K10}
\end{figure}

In this first part, we consider a detonation with a time coefficient $\tau=10^{-1}$ . As observed in Figure
\ref{fig:K10}, the detonation front propagates from left to right with constant maximal values of density, pressure
and velocity, which correspond to the von Neumann state. The slopes for the density, pressure, and velocity, are moderate
on the left side of the shock, {\it i.e.} in the burnt gas zone, but become sharp on the right side, at the interface 
with the unburnt gases. For all the displayed quantities, the curves fit with the reference computation, which has been
performed on a fine grid with $L=14$ scales, {\it i.e.} with $16384$ grid points. 

Table \ref{table:K10} assembles results for computations performed with different maximum number of scales for both the MR and the FV method. 
The relative error is computed on the von Neumann value of the density. 
The numerical value is averaged in time to damp
its oscillations and is compared to the theoretical one. We observe for the MR computations 
a CPU time compression rate growing with increasing number of scales, while the relative error remains comparable with the one of the 
same computation on the regular fine grid. As expected, the error is reduced when increasing the number of scales $L$.

\begin{table}[htbp]
\begin{tabular}{crrrrr}
Method & $L$ & $\epsilon$ & CPU time & CPU compression & relative error \\
\hline 
MR & $9$ & $ 10^{-2}$ & 1 $s$ & 41,34 \% & $1.099 \cdot 10^{-1}$ \\
FV & $9$ & ~ & 3 $s$ & 100,00 \% & $1.109 \cdot 10^{-1}$ \\
MR & $10$ & $5 \cdot 10^{-3}$ & 4 $s$ & 34,28 \% & $6.519 \cdot 10^{-2}$ \\
FV & $10$ & ~ & 13 $s$ & 100,00 \% & $6.601 \cdot 10^{-2}$ \\
MR & $11$ & $2.5 \cdot 10^{-3}$ & 11 $s$ & 21,86 \% & $3.554 \cdot 10^{-2}$ \\
FV & $11$ & ~ & 54 $s$ & 100,00 \% & $3.617 \cdot 10^{-2}$ \\
MR & $12$ & $1.25 \cdot 10^{-3}$ & 40 $s$ & 18,75 \% & $1.740 \cdot 10^{-2}$ \\
FV & $12$ & ~ & 3 $min$ 37 $s$ & 100,00 \% & $1.783 \cdot 10^{-2}$ \\
\end{tabular}
\caption{Chapman-Jouguet detonation front: CPU time, CPU compression and relative error on
the von Neumann density for different scales and for the FV and MR methods, $\tau = 10^{-1}$.}
\label{table:K10}
\end{table}

% -------------------------------------------------------------------------------------------------------------------------
\subsubsection{Influence of the stiffness}

\begin{figure}[htbp]
\includegraphics[width=0.5 \textwidth]{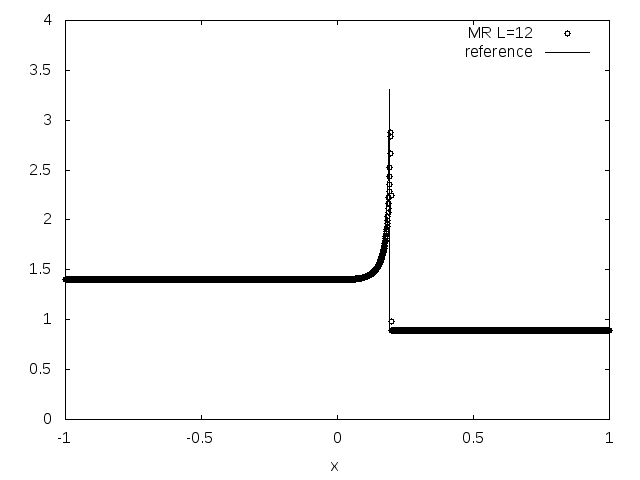}
\includegraphics[width=0.5 \textwidth]{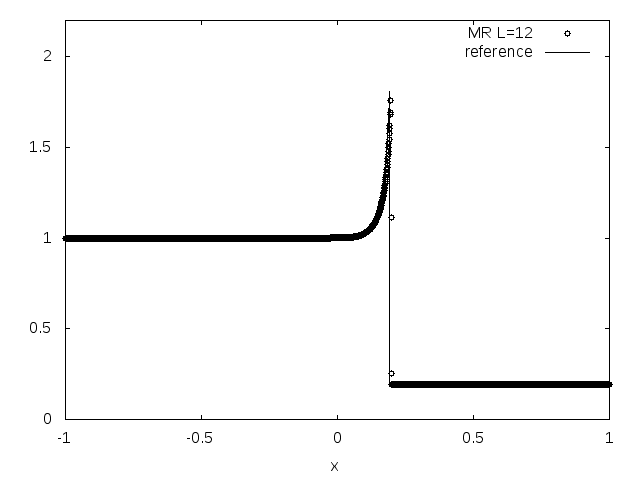}
\\
\includegraphics[width=0.5 \textwidth]{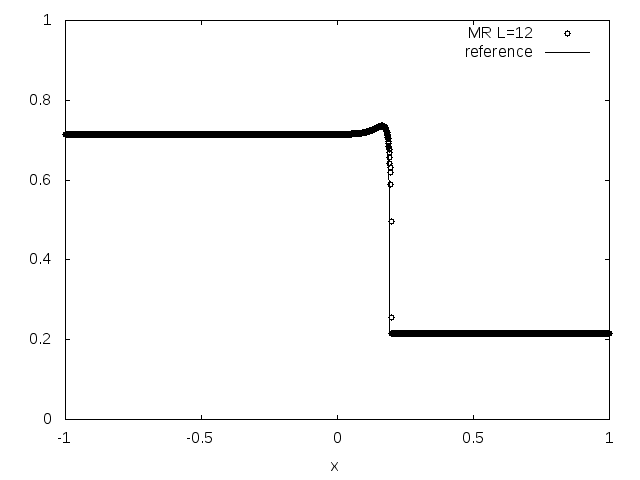}
\includegraphics[width=0.5 \textwidth]{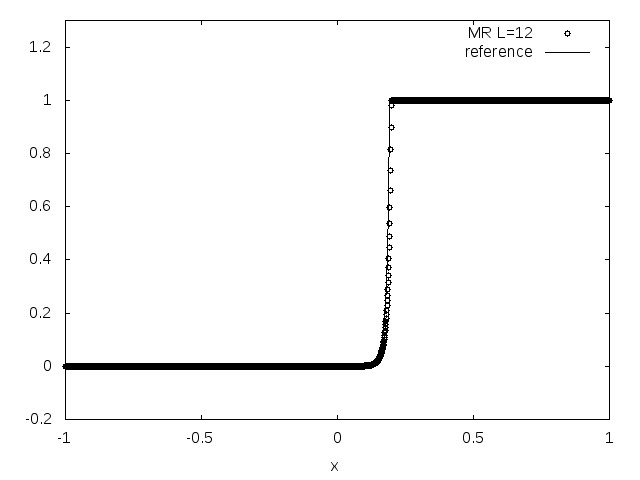}
\\
\includegraphics[width=0.5 \textwidth]{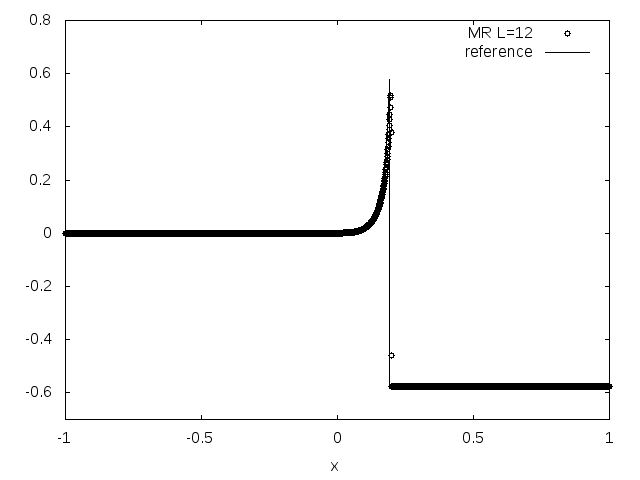}
\includegraphics[width=0.5 \textwidth]{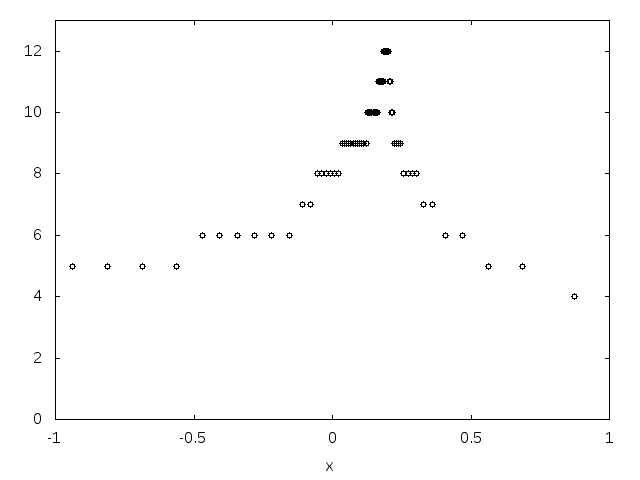}
\caption{Chapman-Jouguet detonation front:
density ({\it top, left}), pressure ({\it top, right}), temperature ({\it middle, left}),
partial mass of the limiting reactant ({\it middle, right}), velocity ({\it bottom, left}) and 
adaptive mesh ({\it bottom, right}) at $t=2.2$, $L=12$ scales, $\epsilon=1.25 \cdot 10^{-3}$, $\tau=10^{-2}$.}
\label{fig:K100}
\end{figure}

\begin{figure}[htbp]
\includegraphics[width=0.5 \textwidth]{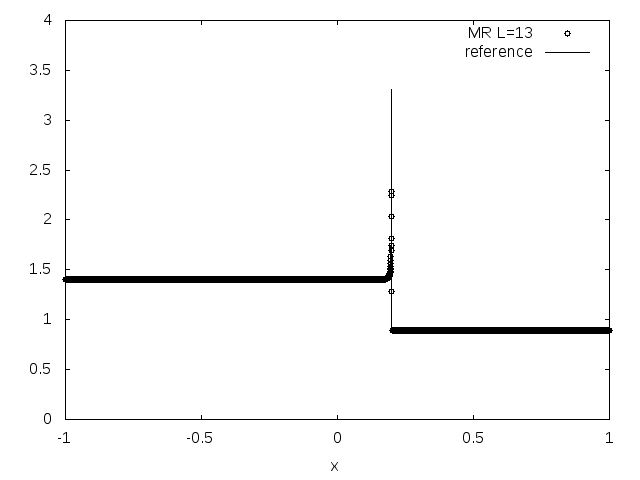}
\includegraphics[width=0.5 \textwidth]{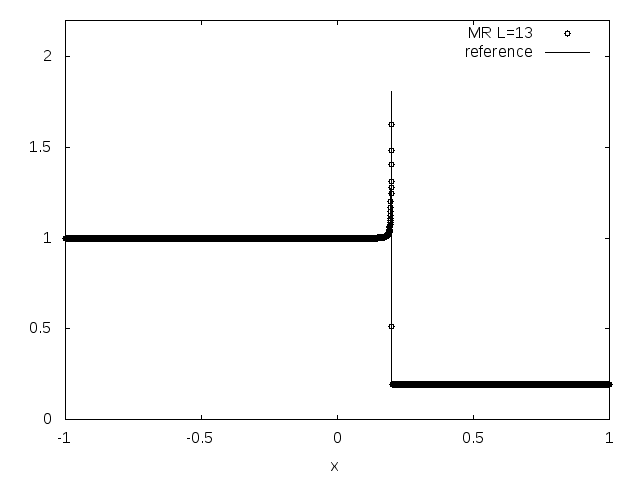}
\\
\includegraphics[width=0.5 \textwidth]{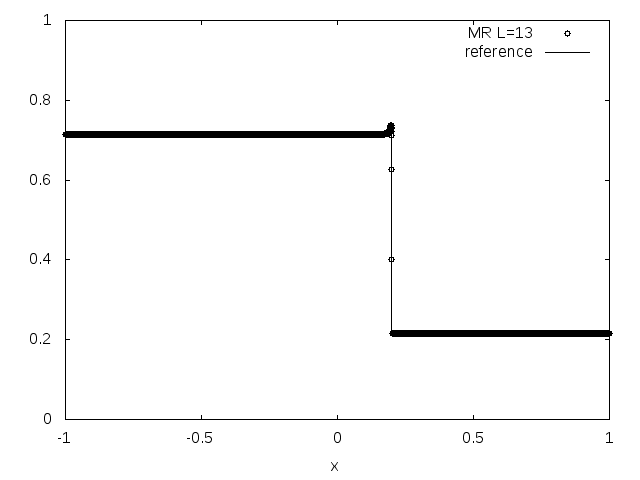}
\includegraphics[width=0.5 \textwidth]{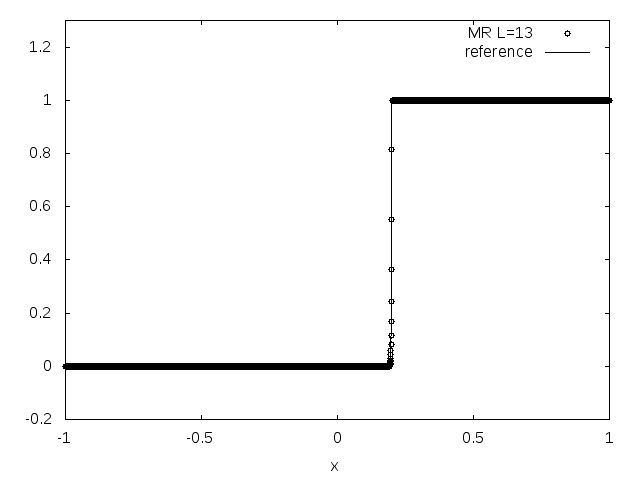}
\\
\includegraphics[width=0.5 \textwidth]{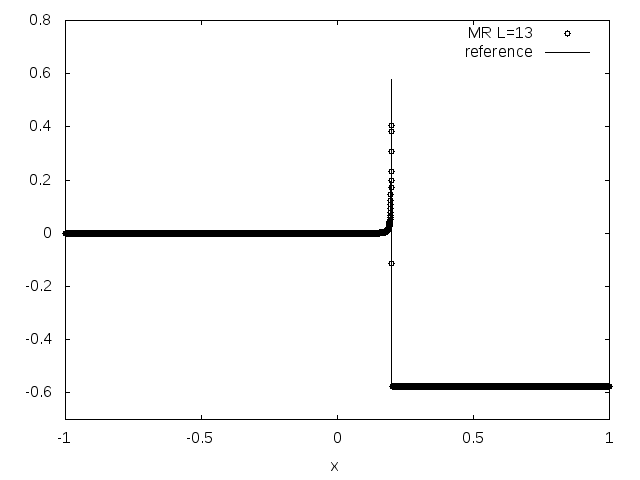}
\includegraphics[width=0.5 \textwidth]{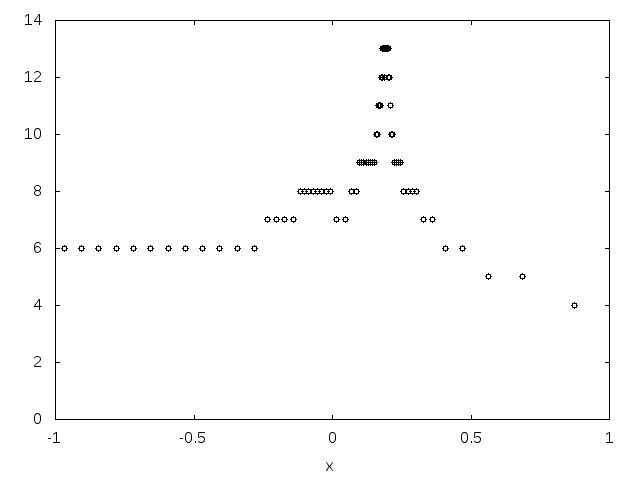}
\caption{Chapman-Jouguet detonation front:
density ({\it top, left}), pressure ({\it top, right}), temperature ({\it middle, left}),
partial mass of the limiting reactant ({\it middle, right}), velocity ({\it bottom, left}) and 
adaptive mesh ({\it bottom, right}) at $t=2.2$, $L=13$ scales, $\epsilon=6.25\cdot 10^{-4}$, $\tau=10^{-3}$.}
\label{fig:K1000}
\end{figure}

In this part, we perform computations of stiff problems and set $\tau$ first to $10^{-2}$, then to $10^{-3}$. 
In order to resolve correctly the detonation front, $13$ scales are required for the case $\tau=10^{-3}$. 
The results are shown in Figures \ref{fig:K100} and \ref{fig:K1000}. 
We observe a much sharper and thinner reaction zone, reduced to a few computational points in
the case $\tau=10^{-3}$. We observe a good agreement with the reference computations, which were computed on a regular fine grid with $L=15$ scales.

Table \ref{table:Kvar} gives the results of the computations for $\tau=10^{-2}$ and $\tau=10^{-3}$. As expected,
a high compression rate is reached when 13 scales are used. But as it is seen in Figure \ref{fig:K1000}, the error
on the von Neumann density remains quite large, even if the detonation front is well tracked.

\begin{table}[htbp]
\begin{tabular}{crrrrrr}
Method & $\tau$ & $L$ & $\epsilon$ & CPU time & CPU comp. & rel. error \\
\hline 
MR & $10^{-2}$ & $12$ & $1.25 \cdot 10^{-3}$ & 43 $s$ &  20,08 \% & $1.336 \cdot 10^{-1}$ \\
FV & $10^{-2}$ & $12$ & ~         & 3 $min$ 37 $s$ & 100,00 \% & $1.342 \cdot 10^{-1}$ \\
MR & $10^{-3}$ & $13$ & $6.25 \cdot 10^{-4}$ & 2 $min$ 03 $s$ &  7,56 \% & $3.057 \cdot 10^{-1}$ \\
FV & $10^{-3}$ & $13$ & ~         & 27 $min$ 13 $s$ & 100,00 \% & $3.057 \cdot 10^{-1}$
\end{tabular}
\caption{Chapman-Jouguet detonation front: CPU time, CPU compression and relative error on
the von Neumann density for different values of the stiffness coefficient and for the FV and MR methods.}
\label{table:Kvar}
\end{table}

% -------------------------------------------------------------------------------------------------------------------------
\subsection{Two-dimensional detonation}

Now we consider a two-dimensional problem.
The interaction between a detonation wave and a pocket of unburnt gas is simulated. 
The computational domain is $[0,4] \times [-1,1]$, the initial planar detonation front is located at $x=1.7$ and the initial spot of unburnt gas is centered in $(x_0,y_0) = (2,0)$. 
The radius of the circular pocket is $r_0 = 0.1$. The parameters of the
detonation front are the same as in the non-stiff case, {\it i.e.} we choose $\tau = 10^{-1}$. 
The dimensionless final time is $t=1.3$.

\begin{figure}[htbp]
\includegraphics[width=0.5 \textwidth]{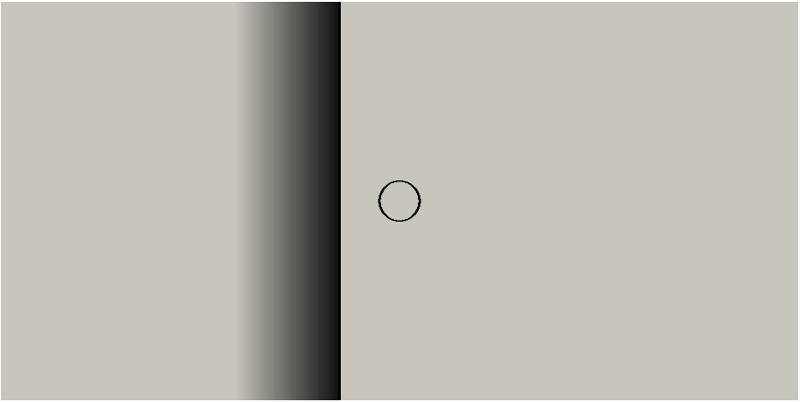}
\includegraphics[width=0.5 \textwidth]{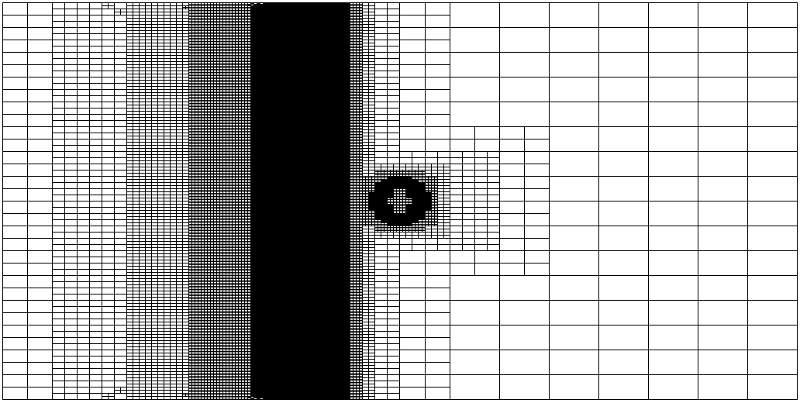}
\\
\includegraphics[width=0.5 \textwidth]{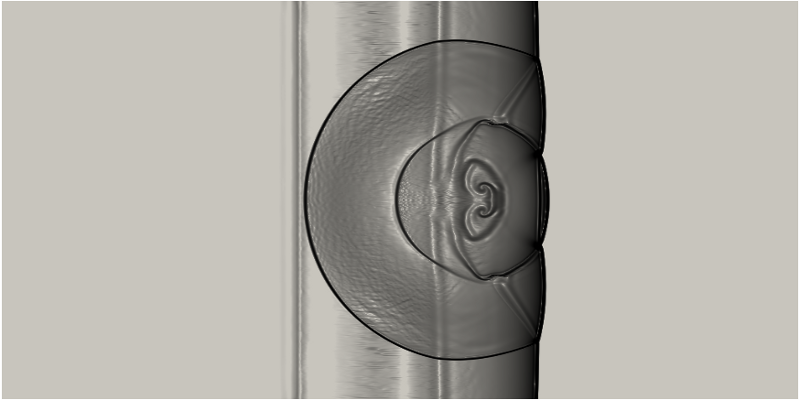}
\includegraphics[width=0.5 \textwidth]{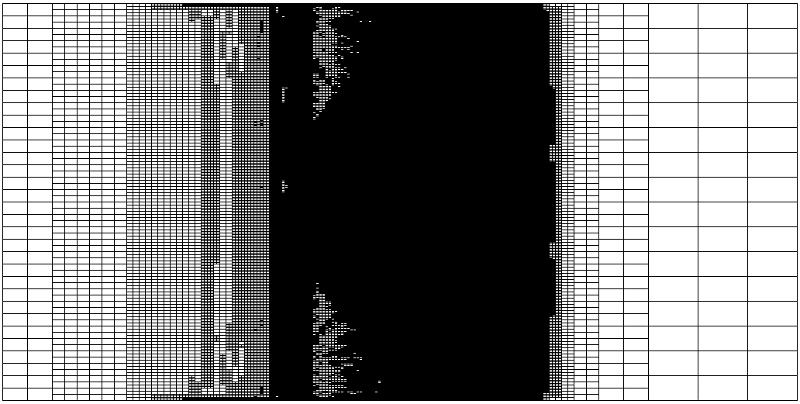}
\\
\includegraphics[width=0.5 \textwidth]{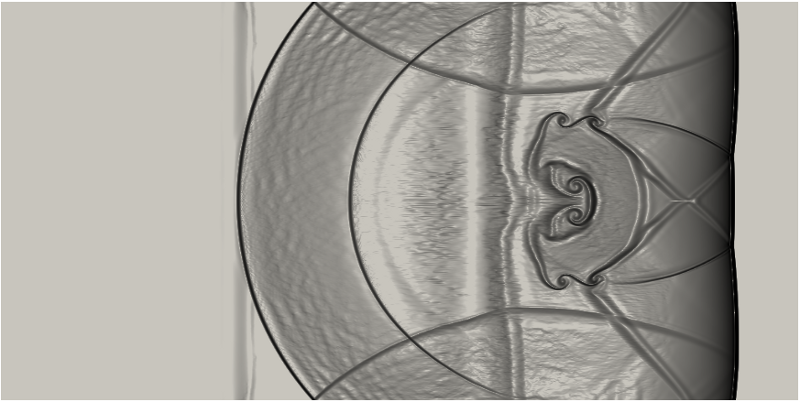}
\includegraphics[width=0.5 \textwidth]{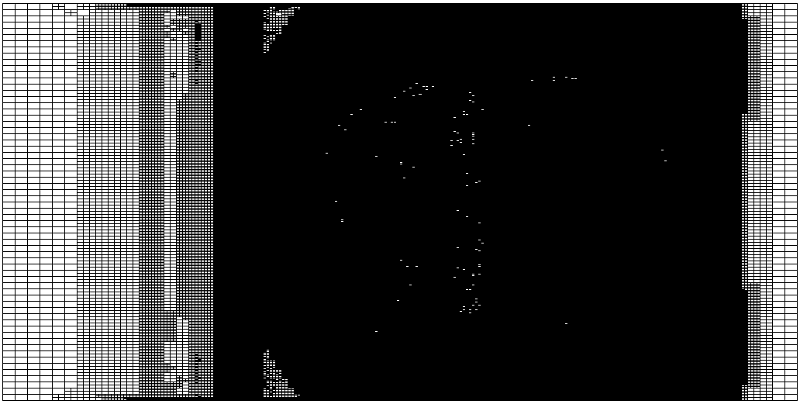}
\caption{Time evolution of the interaction between a detonation wave an a pocket of unburnt gas computed with the adaptive MR method. Numerical Schlieren
of the density gradient ({\it left}) and corresponding meshes ({\it right}) at $t=0$ ({\it top}), $t=0.65$ ({\it middle})
and $t=1.3$ ({\it bottom}).}
\label{fig:2D}
\end{figure}

In Figure \ref{fig:2D}, we observe the destabilization of the planar front by the pocket of unburnt gas. We remark circular
structures due to the expansion of the pocket in the three directions and their interaction with the detonation wave going
from left to right. Recirculations are observed in the center of the pocket, which are advected by the flow. We also
observe reflexions of the waves on the free-slip boundaries in $y=-1$ and $y=1$.

\begin{figure}[htbp]
\includegraphics[width=\textwidth]{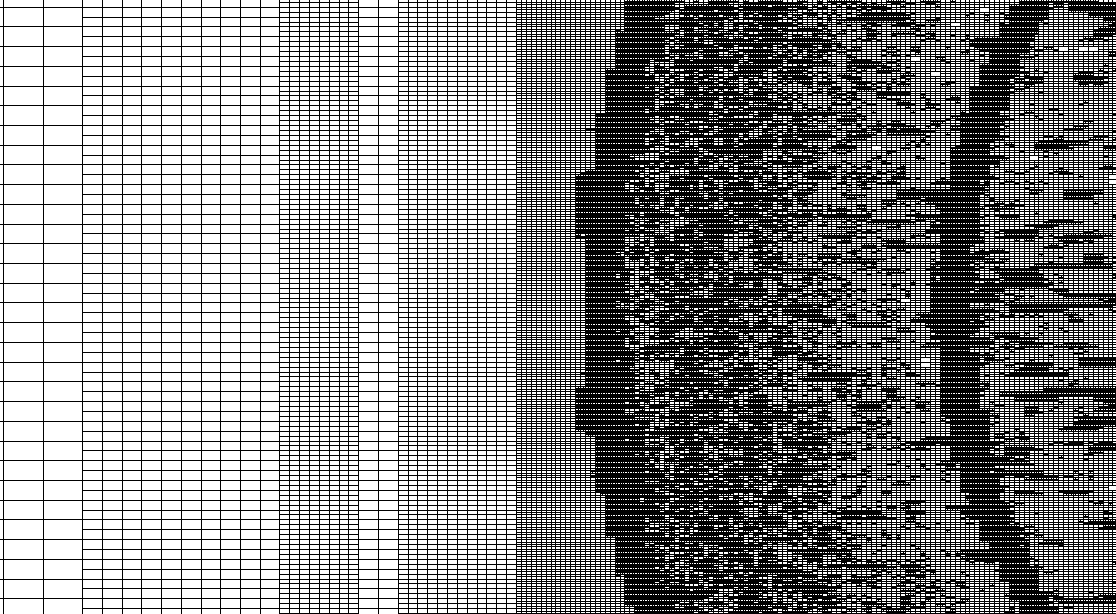}
\caption{Adaptive MR computation of the interaction between a detonation wave an a pocket of unburnt gas: Zoom on the adaptive
mesh at $t=1.3$.}
\label{fig:2Dzoom}
\end{figure}

On the right side of Figure \ref{fig:2D}, we observe that the adaptive mesh follows well the structures of the different
waves that interact in the flow. A zoom in the center of the mesh (Figure \ref{fig:2Dzoom}) enables to see how well the
mesh adapts to all the structures of the flow. The computation, performed on $L=10$ scales {\it i.e.} a maximum of
$2^{2 L} = 1,048,576$ points, requires 2 $h$ 59 $min$ of CPU time on a 16 Core PC. The same
computation on the regular fine grid lasts 14 $h$ 35 $min$. 
Hence the CPU time compression is around 20 \%.

\begin{figure}[htbp]
\includegraphics[width=\textwidth]{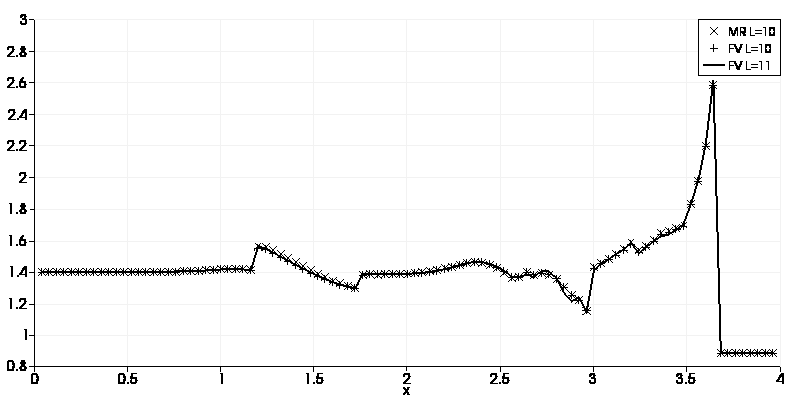}
\caption{Interaction between a detonation wave an a pocket of unburnt gas. One-dimensional cuts of density
for $y=0$ at $t=1.3$. Comparison of the adaptive MR and the FV computations for $L=10$ with the FV reference computation for $L=11$. The total number of grid points in the FV computations is $2^{2 L}$.}
\label{fig:2Dslice}
\end{figure}

Figure \ref{fig:2Dslice} shows a cut of the density at $y=0$. We observe an excellent agreement between the MR and FV
computations with $L=10$ scales, and a good agreement of both curves with the FV reference computation obtained with $L=11$ scales, which validates the grid convergence of the computation. 
{ The total number of grid points in the FV computations is $2^{2 L}$.}

% -------------------------------------------------------------------------------------------------------------------------
\section{Conclusion and perspectives}
% -------------------------------------------------------------------------------------------------------------------------

We presented an extension of the adaptive multiresolution method for the reactive Euler equations, able to deal with
fast chemical reactions.
A finite volume scheme with second order shock capturing schemes is used for space discretization.
Classical Strang splitting is applied to deal with the stiffness of the physical problem in time. 
Space and time adaptivity are then introduced using multiresolution analysis and Runge--Kutta--Fehlberg schemes, respectively. The implementation uses dynamic memory allocation with tree data structures.

Applications to detonation problems in one and two space dimensions validated the algorithm and illustrated the efficiency of the adaption strategy.
The adaptive computations yield both speed-up of CPU time and memory compression with respect to uniform grid computations while the precision is automatically controlled using suitable thresholding techniques.

%Perspective
As perspective we plan the extension of the method to three space dimensions and also to include multi-species chemical reactions {and viscous effects}.
The parallelization of the adaptive method using tree data structures to handle the adaptive grid remains a challenging task for the near future.

% -------------------------------------------------------------------------------------------------------------------------
\section*{Acknowledgements}

This article is dedicated to Professor Henning Bockhorn on the occasion of his 70th birthday, thanking him 
cordially for the great time we had in his group, first in Kaiserslautern and then in Karlsruhe. 
We also thank Margarete Domingues, Ralf Deiterding and Sonia Gomes for constructive
discussions on the topic and fruitful interactions. 
KS and OR thankfully acknowledge financial support from the ANR projects SiCoMHD (ANR-Blanc 2011-045)
and MAPIE (ANR-13-MONU-0002).

% -------------------------------------------------------------------------------------------------------------------------

\end{document}